\documentclass[fleqn,useAMS,usenatbib]{mnras}
\usepackage{mathptmx}
\usepackage{color}
\usepackage{graphics,graphicx,amssymb,amsmath,natbib}
\usepackage{float}
\usepackage{color}
\usepackage{lscape,graphicx}
\usepackage{amsmath}
\usepackage{amssymb}
\usepackage{multirow}
\usepackage{afterpage}
\usepackage{threeparttable}
\usepackage{subfigure}
\usepackage{rotating}
\usepackage{lscape}
\usepackage{caption}
\usepackage{verbatim}
\usepackage{morefloats}
\usepackage[T1]{fontenc}
\usepackage{ae,aecompl}
\usepackage{graphicx}
\usepackage{epstopdf}
\usepackage{amsmath}
\usepackage{amssymb}
\usepackage{enumerate}
\usepackage{dblfloatfix}
\usepackage{lipsum}

\def\asca       {{\em ASCA}\/}
\def\cha    {{\em Chandra}\/}

\def\xmm        {{\em XMM-Newton}\/}

\def\ga         {{CGCG~254-021}\/}

\title[A merger shock in A1367]{A merger shock in Abell 1367}
\author
[Ge et al.]{Chong Ge$^{1}$\thanks{chong.ge@uah.edu}, Ming Sun$^{1}$\thanks{ming.sun@uah.edu}, Ruo-Yu Liu$^{2}$, Lawrence Rudnick$^{3}$, Craig Sarazin$^{4}$,
\newauthor
William Forman$^{5}$, Christine Jones$^{5}$, Hao Chen$^{1}$, Wenhao Liu$^{1}$, Masafumi Yagi$^{6}$,
\newauthor
Alessandro Boselli$^{7}$, Matteo Fossati$^{8}$, Giuseppe Gavazzi$^{9}$\\
$^{1}$Department of Physics and Astronomy, University of Alabama in Huntsville, Huntsville, AL 35899, USA\\
$^{2}$Deutsches Elektronen Synchrotron (DESY), Platanenallee 6, D-15738 Zeuthen, Germany\\
$^{3}$Minnesota Institute for Astrophysics, University of Minnesota, 116 Church Street S.E., Minneapolis, MN 55455, USA\\
$^{4}$Department of Astronomy, University of Virginia, 530 McCormick Rd., Charlottesville, VA 22904, USA\\
$^{5}$Harvard-Smithsonian Center for Astrophysics, 60 Garden Street, Cambridge, MA 02138, USA\\
$^{6}$Optical and Infrared Astronomy Division, National Astronomical Observatory of Japan, Mitaka, Tokyo 181-8588, Japan\\
$^{7}$Aix-Marseille Univ., CNRS, CNES, LAM, Marseille, France\\
$^{8}$Institute for Computational Cosmology and Centre for Extragalactic Astronomy, Department of Physics, Durham University, \\
South Road, Durham DH1 3LE, UK\\
$^{9}$Universit\'a di Milano-Bicocca, piazza della scienza 3, 20100, Milano, Italy\\
}
\begin{document}
\date{Accepted. Received; in original form}

\pubyear{2019}

\maketitle
\begin{abstract}
Multi-wavelength observations show that Abell 1367 (A1367) is a dynamically young cluster, with at least two subclusters merging along the SE-NW direction. With the wide-field \xmm\ mosaic of A1367, we discover a previously unknown merger shock at the NW edge of the cluster. We estimate the shock Mach number from the density and temperature jumps as $M_{\rho}=1.21\pm0.08$ and $M_T=1.60\pm0.07$, respectively. This shock region also corresponds to a radio relic discovered with the {\em VLA} and {\em GBT}, which could be produced by the shock re-acceleration of pre-existing seed relativistic electrons. We suggest that some of the seed relativistic electrons originate from late-type, star-forming galaxies in this region.
\end{abstract}


\section{Introduction} \label{sec:intro}
Galaxy clusters are the most massive virialized systems in the Universe. They form hierarchically through the merger of smaller substructures.
Major cluster mergers are the most energetic events since the Big Bang, with their total kinetic energy reaching $10^{64}$ ergs. Although most of the kinetic energy is in bulk motion of the dark matter, the dominant mass component, a significant portion of this energy is dissipated into the intracluster medium (ICM)
through shock heating. 
The merger shocks provide novel tools to study the ICM on micro and macro scales, where gravity, thermal pressure, magnetic fields, and relativistic particles are at play (e.g. \citealt{2007PhR...443....1M}). 
Although shocks should be pervasive in merging systems, only about twenty shocks in clusters with favorable merger geometry have so far been detected with X-ray observations (e.g. \citealt{2016MNRAS.458..681D}), while more than forty radio relics have been identified in radio (e.g. \citealt{2019SSRv..215...16V}).

A1367 lies at the intersection of two large filaments, which connect to Virgo in the southeast and to Coma in the northeast \citep{2000ApJ...543L..27W}.
A1367 is also a dynamically young cluster based on its high fraction of spiral galaxies, low central galaxy density, and irregular X-ray morphology (e.g. 
\citealt{1983ApJ...265...26B}; \citealt{1984ApJ...285..426B}).
Detailed spatial and dynamical analysis of member galaxies shows a significantly non-Gaussian velocity distribution, and that the cluster is elongated from the southeast (SE) to the northwest (NW) with two main density peaks associated with two substructures \citep{2004A&A...425..429C}.
X-ray observations from \asca\ \citep{1998ApJ...500..138D} and \cha\ \citep{2002ApJ...576..708S} show that the temperature of the NW subcluster ($\sim$4 keV) is higher than the SE subcluster ($\sim$3 keV). Both temperature maps show that the hottest region is near the NW edge.

Here, based on the \xmm\ observations of A1367 we report the detection of a merger shock in the NW region of the cluster. We assume a cosmology with $H_0$ = 70 km s$^{-1}$ Mpc$^{-1}$, $\Omega_m=0.3$, and $\Omega_{\Lambda}= 0.7$. At the A1367 redshift of $z=0.022$, $1^{\prime\prime}=0.445$ kpc.

\section{Data analysis}
\label{s:obs}
Table~\ref{t:obs} lists the detailed information for the \xmm\ observations.
We processed the \xmm\ MOS and pn data using the Extended Source Analysis Software (ESAS), as integrated into the \xmm\ Science Analysis System (SAS; version 17.0.0), following the procedures in \cite{2019MNRAS.484.1946G}.
Our background analysis follows the standard ESAS procedure, aided by the source-free regions in the \xmm\ data and the off-cluster RASS spectrum.
The spectra from MOS/pn and multiple observations are fitted jointly with the XSPEC package. We use the AtomDB (version 3.0.8) database of atomic data and the solar abundance tables from \cite{2009ARA&A..47..481A}. The Galactic column density $N_{\rm H}=1.91\times10^{20}\ {\rm cm}^{-2}$ is taken from the NHtot tool \citep{2013MNRAS.431..394W}.

\begin{table}
 \centering
  \caption{\xmm\ observations}
\tabcolsep=0.1cm  
  \begin{tabular}{@{}lccc@{}}
\hline\hline
Obs-ID & PI & Exp (ks)$^a$ & Clean Exp (ks)$^a$ \\ 
\hline
0005210101 & R. Fusco-Femiano & 32.9/32.9/28.5 & 26.0/26.1/14.9\\
0061740101 & W. Forman & 32.6/32.6/28.0 & 28.7/29.5/17.4\\   
0301900601 & A. Wolter & 29.6/29.6/27.9 & 19.5/19.6/13.0\\
0602200101 & A. Finoguenov & 24.6/24.6/20.7 & 22.5/23.6/14.0\\
0602200201 & A. Finoguenov & 13.6/13.6/9.7 & 12.6/12.3/7.0\\ 
0602200301 & A. Finoguenov & 22.0/22.1/19.5 & 5.4/6.0/0.8\\ 
0823200101 & M. Sun & 71.6/71.6/69.8 & 66.0/66.8/50.4\\ 
\hline
\end{tabular}
\begin{tablenotes}
\item
$^a$EPIC exposures of the MOS1/MOS2/pn cameras.
\end{tablenotes}
\label{t:obs}
\end{table}

\begin{figure*}
\centering
\includegraphics[width=0.44\textwidth,keepaspectratio=true,clip=true]{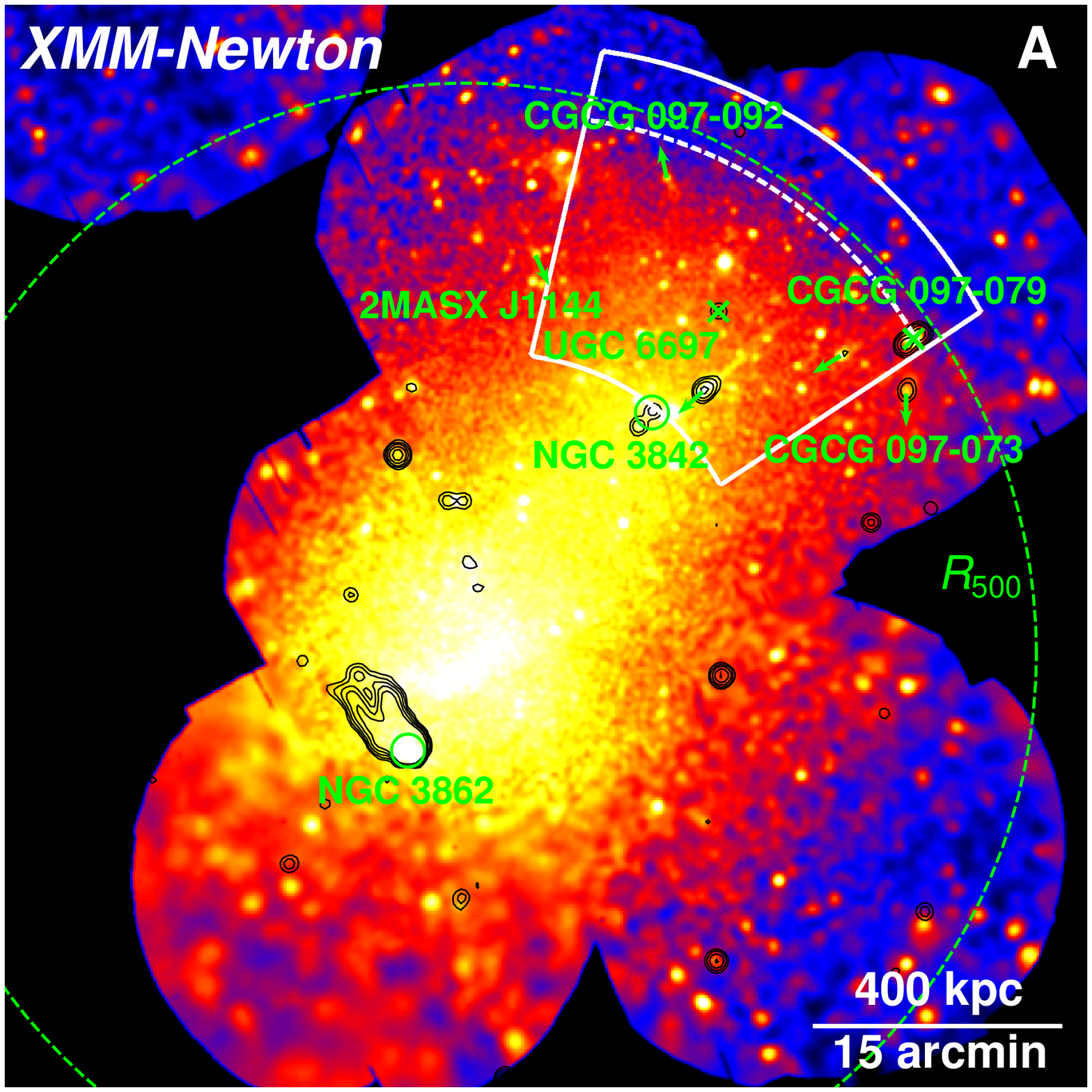}
\includegraphics[width=0.52\textwidth,keepaspectratio=true,clip=true]{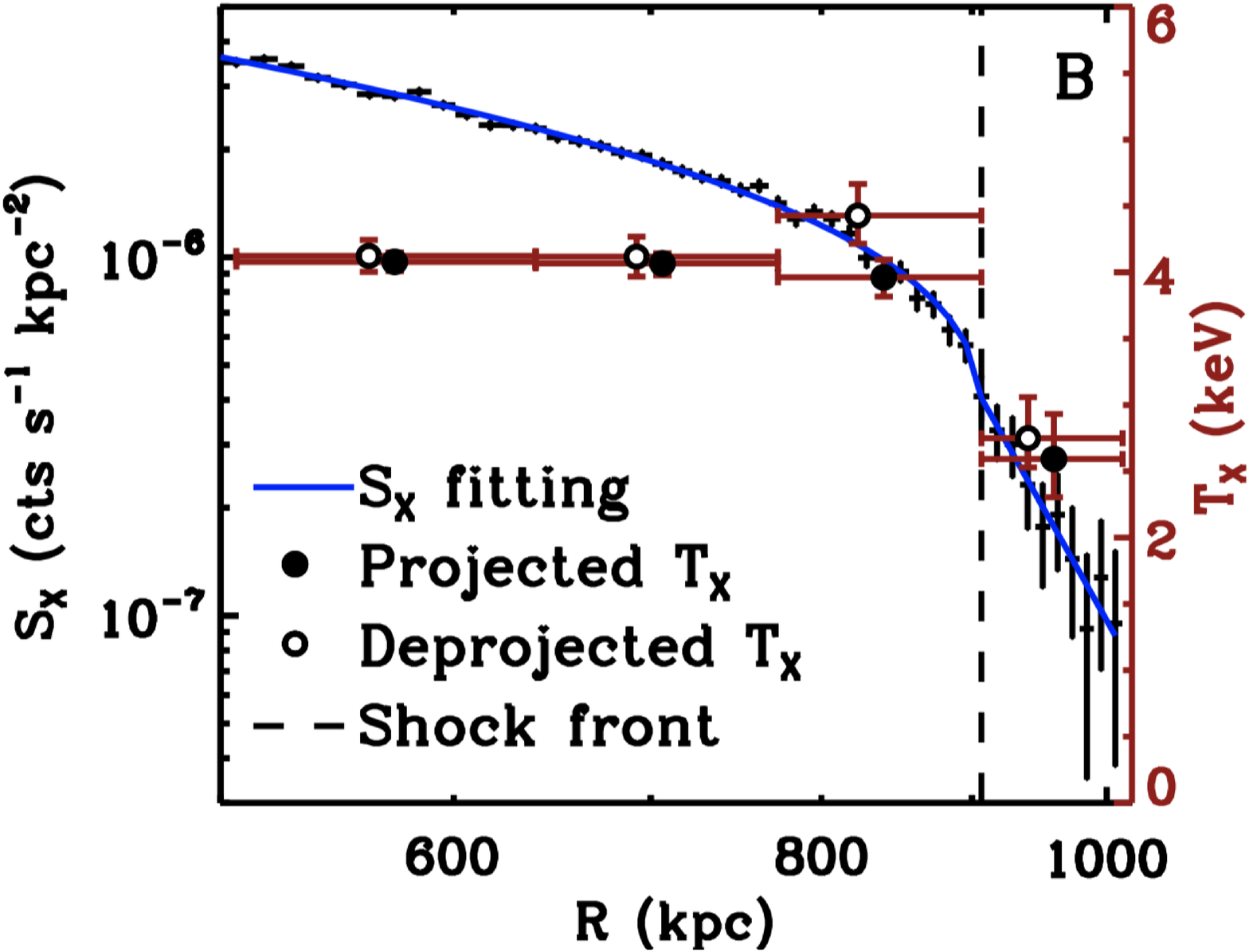}
\includegraphics[width=0.48\textwidth,keepaspectratio=true,clip=true]{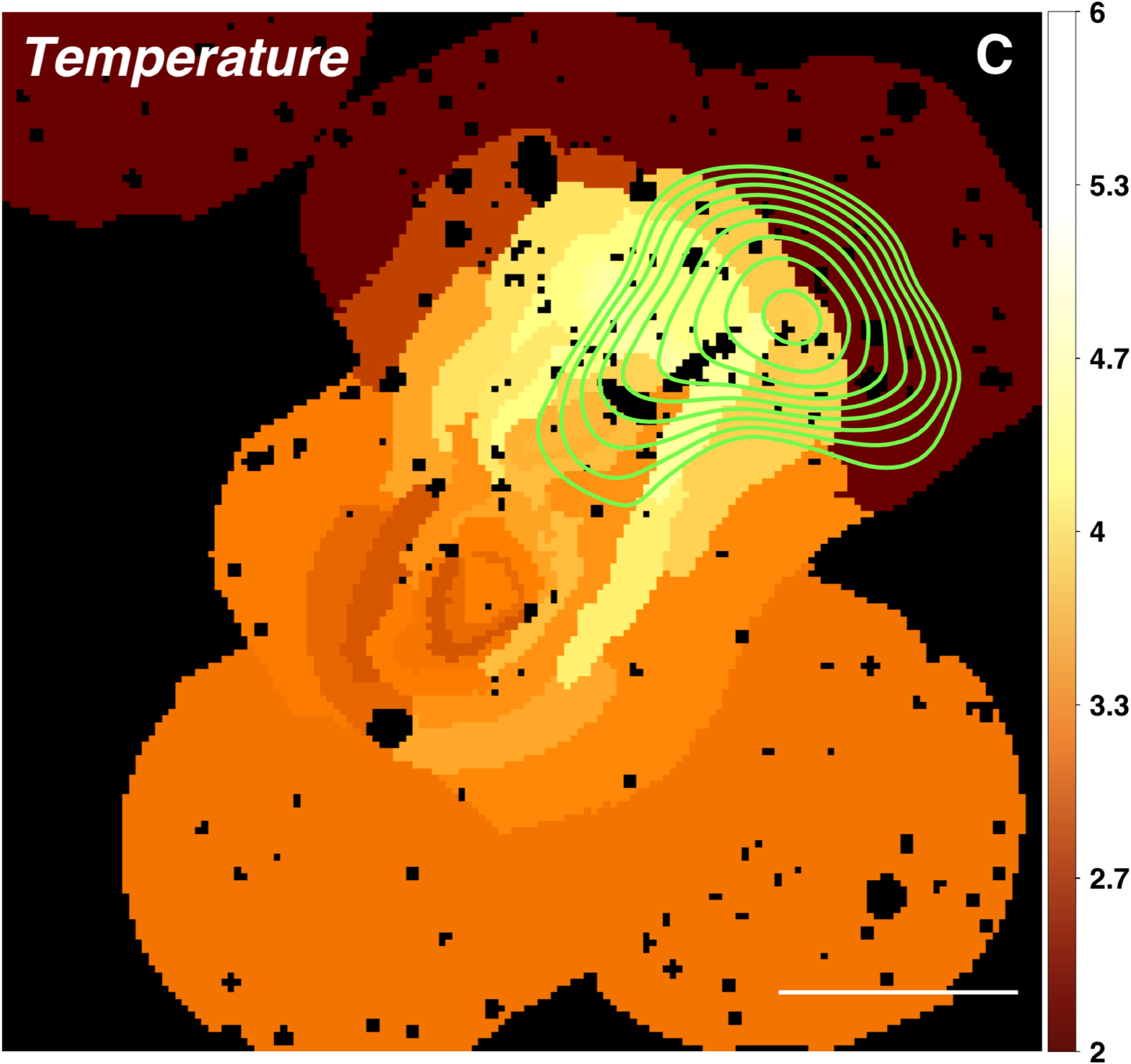}
\includegraphics[width=0.48\textwidth,keepaspectratio=true,clip=true]{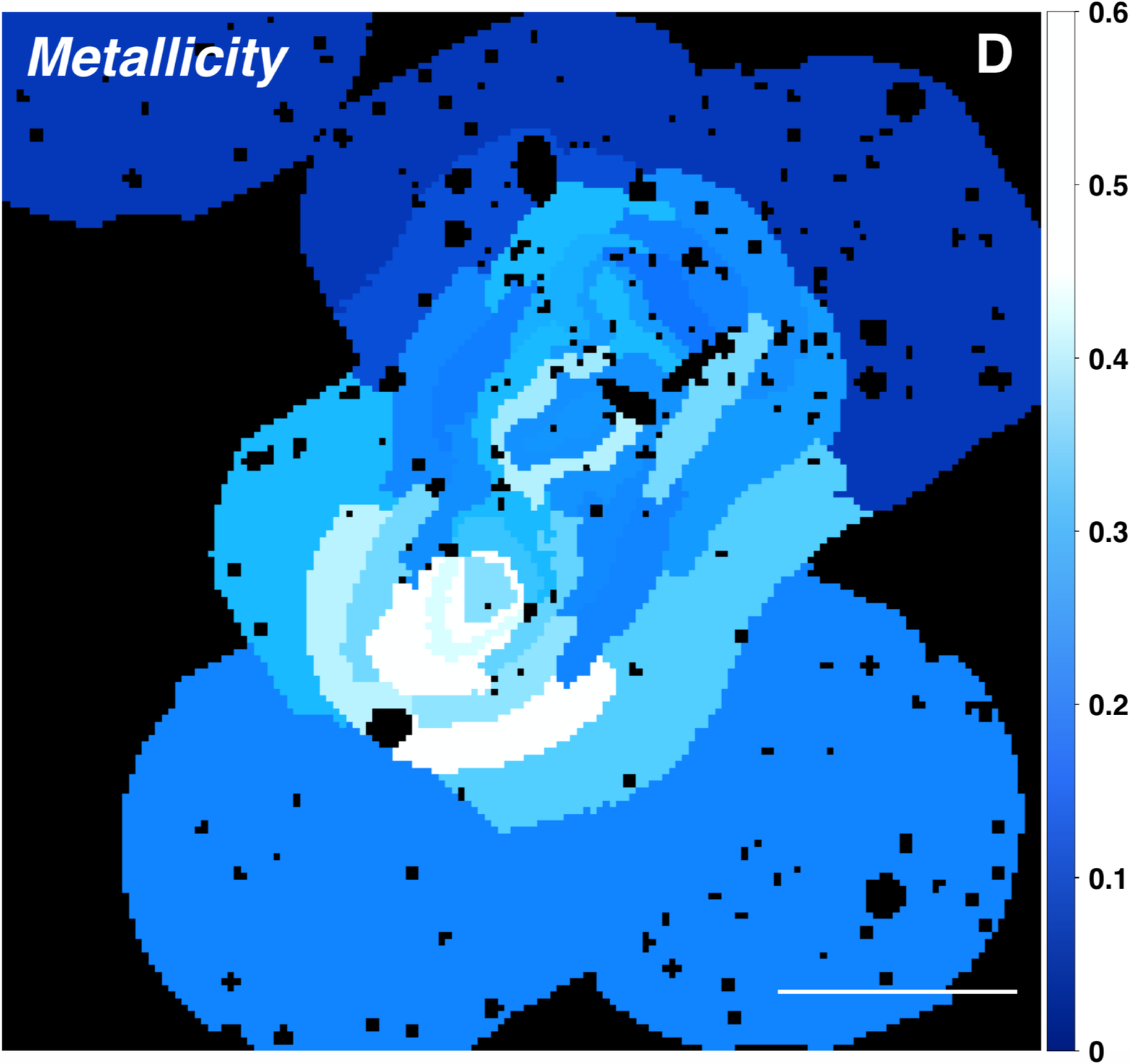}
	\caption{{\it (A)}:
	\xmm\ EPIC combined 0.5-2.0 keV background subtracted, exposure corrected, and smoothed image. Notable features are marked: circles mark the BCGs of the NW (NGC 3842) and SE (NGC 3862/3C 264) subclusters, which are bright X-ray sources; arrows mark the direction of motion for galaxies with $\rm {H}\alpha$ tails emission \citep{2017ApJ...839...65Y}; crosses mark the background radio sources in the post-shock region; black contours show the 1.4 GHz continuum from {\em NVSS} at 4, 8, 16, 32, and 64 mJy beam$^{-1}$; the white sector marks the extraction region for SBP shown in the (B) panel and the dashed line shows a surface brightness edge (shock front); the green dashed circle is $R_{500}=924$ kpc; the bar at bottom right shows 15 arcmin/400 kpc (same for the bottom two panels).
	{\it (B)}: SBP and temperature profiles. The blue line shows the best fitting power-law model with an ellipsoidal emissivity discontinuity. Black dots give the projected temperature, while hollow dots are the deprojected temperature. The dashed line marks the shock front.  
	{\it (C)}: Gas temperature map with a color bar unit of keV, overlaid with contours from the 1.4 GHz radio map of diffuse emission from {\em GBT} \citep{2013ApJ...779..189F}. 
	{\it (D)}: ICM metallicity map with a unit of $Z_\odot$.}
	 \label{fig:multi}
\end{figure*}

\section{results}
\subsection{Merger shock}
Fig.~\ref{fig:multi}(A) shows the 0.5-2.0 keV image combined from all seven \xmm\ observations. As seen in previous X-ray observations, the diffuse X-ray emission is elongated in the SE to NW direction. We note a brightness edge in the NW, which is marked with a white dashed line. Our further analysis suggests that this is a shock front.

We extract a surface brightness profile (SBP) from the NW subcluster in self-similar elliptical pie annuli enclosed within the white sector shown in Fig.~\ref{fig:multi}(A). The center of these annuli is on the X-ray peak. The shape of the ellipses matches approximately the sharp outer surface brightness edge. Fig.~\ref{fig:multi}(B) shows the SBP; the dashed line marks the location of the surface brightness edge, where there is an emissivity jump and the slope changes.
To fit the SBP, we assume the X-ray emissivity is constant on self-similar ellipsoidal surfaces, the principal axis of these ellipsoids is in the plane of the sky, and the emissivity has an ellipsoidal discontinuity and follows different power-law functions of the elliptical radius inside and outside this discontinuity (\citealt{2016arXiv160607433S}).
The best-fit gas density jump across the edge is $\rho_2/\rho_1=1.30\pm0.12$, where suffixes 2 and 1 denote the post-shock and pre-shock regions respectively. We then apply the Rankine-Hugoniot jump condition (
$\gamma=5/3$ for a monoatomic gas) to determine the shock Mach number of $M_{\rho}=1.21\pm0.08$.

We also measure the temperature jump from X-ray spectroscopy in the post-shock and pre-shock regions. Fig.~\ref{fig:multi}(B) shows the projected temperature profile, taken from the same sector as used for the SBP. The post-shock projected $T_X$ is $\sim 4.0$ keV, while the pre-shock projected $T_X$ is $\sim 2.5$ keV. To deproject the temperature, we use a similar method to the ``onion-peeling'' technique \citep{2003ApJ...598..250S},
with the assumptions of spherical symmetry and uniform cluster temperature in each spherical shell. We use Monte Carlo simulations to populate counts in a 3D shell, and then project them on a 2D image where each pixel represents the number of counts integrated along the line of sight in the shell. The normalizing ratio in different temperature extraction sectors is estimated from the counts ratio in the projected 2D image after applying the same sector mask, where the point sources, chip gaps, and wedge shaped regions are accounted for. However, the pre-shock region is close to $R_{500}$ and the outside region has too few counts for spectral fitting. Instead, we do a model extrapolation, assuming the density follows a $\beta$-model ($\beta=2/3$) and use the temperature profile from \cite{2006ApJ...640..691V}. The resulting deprojected $T_X$ profile is presented in Fig.~\ref{fig:multi}(B).  The deprojected $T_X$ jump is $T_2/T_1=1.60\pm0.07$. Applying the Rankine-Hugoniot jump condition, the corresponding Mach number is $M_T=1.60\pm0.07$.
 
Both the gas density jump of $1.30\pm0.12$ and temperature jump of $1.60\pm0.07$ indicate a robust shock in the NW region of A1367. Although there is a small discrepancy between the Mach numbers $M_{\rho}$ and $M_T$, which may be the result of the assumptions we used, e.g. ellipsoidal or spherical symmetry, the shock propagation is in the plane of the sky or the result of projection problems, the uncertain pre-shock temperature from model extrapolation.

\subsection{Temperature and metallicity maps}
We apply the Contour Binning algorithm \citep{2006MNRAS.371..829S} to generate spatial bins that closely follow the X-ray surface brightness. After masking the point sources and some extended features, a signal-to-noise ratio of 100 is selected, which requires $\sim 10,000$ background-subtracted counts per bin. A geometric constraint value of 2 is also selected to ensure the bin region would not be too elongated. We then extract the spectra and response files from individual bins. The best fitting temperature and metallicity maps are shown in Fig.~\ref{fig:multi}.

The \xmm\ temperature map is consistent with previous ones from \asca\ \citep{1998ApJ...500..138D} and \cha\ \citep{2002ApJ...576..708S}. In general, the NW subcluster is hotter than the SE one. The hottest ICM is located in the post-shock region near the NW edge. Beyond this edge, the temperature suddenly drops; this edge is a typical feature of a shock front. The metallicity map shows that the SE subcluster has higher abundances than the NW one.
The metallicity peak of $\sim 0.5\ Z_\odot$ is located near the core of the SE subcluster. The temperature is also lower in this region, which could indicate a cool core.

\section{Discussion}
\subsection{Diffuse radio emission}
The NW region of A1367 has been known to host a radio relic for over 40 years \citep{1978A&A....69..355G,1987A&A...186L...1G,2013ApJ...779..189F}.
We overlay the diffuse radio contours from the {\em GBT-NVSS} residual map \citep{2013ApJ...779..189F} on the gas temperature map in Fig.~\ref{fig:multi}(C).
The relic's largest linear scale (LLS) is $\sim 600$ kpc with a 1.4 GHz flux of $S_{1.4}=232\pm28$ mJy (note $S_{1.4}=103\pm21$ mJy from \citealt{1978A&A....69..355G}). Its peak (total) fractional polarization is $\sim$18\% (15\%) \citep{2013ApJ...779..189F}, and the high level of polarization ($\gtrsim$ 10\% at GHz) is typical for a radio relic. 
The relic is coincident with the shock region. With a radio luminosity of $P_{\rm 1.4}=2.5 \times 10^{23}\ {\rm W\ Hz}^{-1}$ and a mass of $M_{500}=2.3 \times 10^{14}\ M_{\odot}$, estimated from the mean X-ray temperature \citep{2009ApJ...693.1142S}, A1367's radio relic is among the faintest known with the ICM shock detected and still on the average mass-luminosity relation for relics (e.g. \citealt{2014MNRAS.444.3130D}). The origin of radio relics in galaxy clusters is still under debate. Theoretical work suggests that the particle acceleration efficiency from the thermal pool of the ICM to the synchrotron emitting electrons via the diffusive shock acceleration (DSA) mechanism is very low (e.g. \citealt{2011ApJ...734...18K}), especially for shocks with $M<3$. Observations from the Bullet cluster show that the diffuse radio features, associated with shocks of similar Mach numbers, are significantly different from each other \citep{2015MNRAS.449.1486S}. A possible explanation is that pre-existing seed relativistic electrons are necessary for the shock to re-accelerate or compress. The seed electrons are typically proposed to arise from structure formation shocks (e.g.  \citealt{2013MNRAS.435.1061P}) or radio galaxies with an active galactic nucleus (AGN), which is either currently active or was active in the past (e.g. \citealt{2017NatAs...1E...5V}). Fig.~\ref{fig:multi}(A) shows the 1.4 GHz radio continuum contours from {\em NVSS}. Excluding the background radio AGN, we do not find any bright radio AGN close to the shock region, except for the small scale ($\sim$ 20 kpc) radio emission associated with the brightest cluster galaxy (BCG) NGC~3842.

On the other hand, we note there are three starburst galaxies (CGCG~097-073, 097-079, and 097-087/ UGC~6697) with ${\rm H}{\alpha}$ and radio tails (\citealt{1987A&A...186L...1G}; \citealt{1995A&A...304..325G}; \citealt{2017ApJ...839...65Y}) close to the shock region. Two of them are brighter than NGC~3842 at 1.4 GHz (12.3 mJy from {\em NVSS}): UGC 6697 (53.9 mJy), CGCG~097-073 (21.5 mJy), CGCG~097-079 (5.0 mJy). 
We estimate their star formation rate (SFR) from the {\em GALEX} FUV and {\em WISE} 22 $\mu$m data with the relation derived by \citet{2011ApJ...741..124H}: UGC~6697 --- $5.1\,M_\odot \rm yr^{-1}$, CGCG~097-073 --- $1.0\,M_\odot \rm yr^{-1}$, CGCG~097-079 --- $2.1\,M_\odot \rm yr^{-1}$.
All three also have X-ray tails opposite to their direction of motion as marked in Fig.~\ref{fig:multi}(A). The ram pressure stripping could strip the gas from the galaxies, including the relativistic electrons. Star formation also can proceed in the stripped gas in the intracluster space \citep[e.g.][]{2010ApJ...708..946S}.
Indeed, the 1.4 GHz image from \cite{1987A&A...186L...1G} suggests that the radio relic is fed by the radio tails of CGCG~097-073 and CGCG~097-079, while the radio tail of UGC~6697 also points in this direction.
Thus, in addition to the AGN, star formation activities, e.g. acceleration of relativistic electrons in the supernova remnant (SNR), can potentially provide seed relativistic electrons in the ICM.
Meanwhile, the 1.4 GHz radio luminosity function in nearby clusters \citep{2002AJ....124.2453M} shows that the AGNs dominate at the high luminosities, while the star-forming (SF) galaxies dominate at the low luminosities. Alhough the total radio luminosity is higher from the AGNs, they are more concentrated towards the cluster center than SF galaxies. The contribution of cosmic-ray electrons (CRe) from SF galaxies could be significant at the cluster outskirts, where radio relics are usually found, especially for a cluster accreting a galaxy group with a higher fraction of SF galaxies.  
The possibility of CRe from intra-cluster SNe producing the radio mini-halos has also been discussed \citep{2019MNRAS.tmpL..26O}.
Since the lifetime of fossil electrons stripped from these SF galaxies is $\sim$ Gyr, they have lost most of their energy and do not radiate within the observable radio band beyond the radio continuum tails. However, these fossil electrons could be efficiently
re-accelerated at shocks and are therefore able to create bright radio relics (e.g. \citealt{2017NatAs...1E...5V}). 

Based on the assumption that all the radiating electrons originate from CRe injected in SF galaxies, we make a rough estimate to the SFR needed to supply the observed relic luminosity from shock re-acceleration of these CRe. We start with the estimate of the total number of re-accelerated electrons required from the radio data.
The typical frequency $\nu_c$ of synchrotron radiation by an electron of energy $E_e$ is given by
$\nu_c = 1.4\left({E_e}/{10\rm GeV}\right)^2\left({B}/{3\mu\rm G}\right) \,\rm GHz$.
The synchrotron radiation power of an electron that emits 1.4\,GHz photon is $P_{\rm syn}=3.8\times 10^{-18} (\nu/{1.4\rm GHz})(B/{3\mu\rm G})\,\rm erg~s^{-1}$.
The observed flux is $S_{1.4} = 232\rm mJy$, corresponding to a luminosity of $\nu L_\nu=3.5\times 10^{39}\rm erg~s^{-1}$ given a luminosity distance of 95\,Mpc for the cluster. This implies that the total number of 1.4\,GHz-radiating electrons is $N_{e, \rm 1.4GHz}\simeq 10^{57}(\nu/{1.4\rm GHz})^{-1}(B/3\mu \rm G)^{-1}$. 
Assuming that the merger shock (MS) accelerates seed electrons to a power-law spectrum of $dN_{e,\rm MS}/dE_e=K_{\rm MS}E_e^{-2}$ as anticipated in the canonical DSA theory, with $K_{\rm MS}$ being the normalization constant, the total number of re-accelerated electrons is
\begin{equation}\label{eq:NMS}
N_{e, \rm MS}^{\rm tot}=\int \frac{dN_{e,\rm MS}}{dE_e}dE_e \simeq K_{\rm MS}E_{e,\rm min}^{-1}
\end{equation}
where $E_{e,\rm min}$ is the minimum electron energy in the CRe distribution.
The electron number can also be given by $N_{e,\rm 1.4GHz}\simeq K_{\rm MS}(10{\rm GeV})^{-1}(\nu/1.4{\rm GHz})^{-1/2}(B/3{\mu \rm G})^{1/2}$. Substituting the expression of $K_{\rm MS}$ obtained from the above two $N_{e,\rm 1.4GHz}$ expressions to Eq.~(\ref{eq:NMS}), 
we then find that the total number of re-accelerated electrons in the merger shock is
$N_{e,\rm MS}^{\rm tot}={10^{57}}({E_{e,\rm min}}/{{10\rm \,GeV}})^{-1}({\nu}/{{1.4\rm GHz}})^{-1/2}({B}/{3\mu {\rm G}})^{-3/2}$.
Considering a time interval of 1\,Gyr for the system, the required CRe injection rate is $N_{e,\rm MS}^{\rm tot}/1\rm \,Gyr$. Due to the conservation of the total electron number between that at injection and that at re-acceleration, we can translate the CRe injection rate to SFR rate as follows. We denote the total luminosity of CRe (i.e., the energy injection rate of CRe) of all the relevant SF galaxies by $L_{e,\rm SF}$, and assume that these electrons also follow a power-law distribution of $d\dot{N}_{e,\rm SF}/dE_e=K_{\rm SF}E_e^{-2}$. We find
$\dot{N}_{e,\rm SF}={L_{e,\rm SF}}E_{e,\rm min}^{-1}/\ln(E_{e,\rm max}/E_{e,\rm min})$,
where $\ln(E_{e,\rm max}/E_{e,\rm min})\sim 10$. Thus, to supply sufficient seed electrons for the re-acceleration, the required CRe injection luminosity $L_{e,\rm SF}$ can be found by equating $\dot{N}_{e,\rm SF}$ to $N_{e,\rm MS}^{\rm tot}/1\rm \,Gyr$, 
\begin{equation}
\begin{split}
L_{e,\rm SF}\simeq &5\times 10^{39} \left(\frac{\nu L_\nu}{3.5\times 10^{39}\rm erg~s^{-1}} \right)\left(\frac{t}{1\,\rm Gyr}\right)^{-1}\\
&\times \left(\frac{\nu}{1.4\rm GHz}\right)^{-1/2}\left(\frac{B}{3\mu\rm G}\right)^{-3/2} \rm erg~s^{-1}.
\end{split}
\end{equation}
In a galaxy, CRe are usually believed to be produced by SNRs or possibly powerful winds of massive stars, both of which are related to the star formation process. Given that the CRe injection luminosity is $\sim 10^{39}\rm erg~s^{-1}$ \citep{2010ApJ...722L..58S} while the SFR is $\sim 1\,M_\odot \rm yr^{-1}$ in our Galaxy, we estimate the SFR needed to support such a CRe injection luminosity to be $\sim 5\,M_\odot \rm yr^{-1}$. 
We therefore conclude that the starburst galaxies in A1367 can provide a significant number of seed relativistic electrons through stripping.

\subsection{Merger dynamics}
A1367 is undergoing a merger along the NW-SE  filament towards Virgo. We note that the gas temperature and metallicity are higher in the SW  than in the NE (Fig.~\ref{fig:multi}), which could be due to the accretion of cooler, lower metallicity gas along the filament towards Coma in the NE \citep{2003astro.ph..1476F}. 

Fig.~\ref{fig:multi}(A) shows that the two BCGs in the subclusters are offset from the X-ray peak. This is very similar to the Bullet cluster \citep{2004ApJ...606..819M}, where the collisionless dark matter and galaxies have passed through each other, and the collisional hot gas lags behind. The X-ray peak is in the SE subcluster. Its cool and ridge-like structure \citep{2002ApJ...576..708S} may indicate that it is a disturbed cool core with a cold front \citep{2010A&A...516A..32G}. 
In the NW, the density and temperature jumps show the presence of a shock.
The sound speed is $c_s = 840\ {\rm km\ s}^{-1}$ in the pre-shock region from its temperature of $T_1=2.8$ keV. The Mach number of $M_T=1.60$ (the temperature jump
should be less sensitive to projection effects) gives a shock velocity of $v=Mc_s=1344\ {\rm km\ s}^{-1}$. The distance between the BCGs of the merging subclusters is 680 kpc. Assuming that the subcluster moves at the shock velocity  and the merger in the plane of the sky, the two subclusters passed through each other 0.7 Gyr ago.

In addition to the merger between the NW and SE subclusters, \cite{2004A&A...425..429C} found that three groups with a higher fraction of SF galaxies are falling into A1367. \cite{2010MNRAS.403.1175S} found that A1367 is over-populated with late-type galaxies, and the relatively gas rich blue galaxies are predominantly found in a region that is coincident with the radio relic (Fig.~14 in \citealt{2010MNRAS.403.1175S}). The cluster merger and infalling groups would change the environment of member galaxies. The ram pressure stripping and tidal interaction could quench their star formation and transform late-type galaxies to early-type galaxies (e.g. \citealt{1985ApJ...292..404G}; \citealt{2006PASP..118..517B}). The accelerated galaxy evolution and rapid transformation indicate a short period of enhanced star formation (e.g. \citealt{2005AJ....129...31O}). Thus the sudden change in the environment of late-type galaxies due to the cluster merger could trigger star formation activity, which would produce the seed electrons for the shock re-acceleration, and then produce the radio relic as observed in A1367. 

\vspace{-17pt}
\section{Summary}
We are witnessing the assembling of A1367 at a node of the cosmic web, where the merging of subclusters and accretion of groups are taking place. The two large subclusters are merging along the SE-NW direction. We find a merger shock at the NW edge of the cluster. The two subclusters passed through each other $\sim$ 0.7 Gyr ago, based on the separation of the two subcluster BCGs and the shock velocity. There is a radio relic coincident with the shock region. We propose that the bulk of the seed relativistic electrons for shock re-acceleration to produce the radio relic in A1367 come from late-type, SF galaxies in this region. As star formation can provide seed relativistic electrons, we expect mergers in spiral-rich galaxy groups or proto-clusters (when the cosmic SFR was much higher) to also generate radio relics in these systems.

\section*{Acknowledgements}

Support for this work was provided by the NASA grant 80NSSC18K0606 and the NSF grant 1714764. Partial support for LR comes from NSF grant AST 17-14205 to the University of Minnesota. MF has received funding from the European Research Council (ERC) under the European Union's Horizon 2020 research and innovation programme (grant agreement No 757535).


\end{document}